\documentclass[10pt,english,aps,a4paper]{revtex4}
\usepackage{ae,aecompl}
\usepackage[T1]{fontenc}
\usepackage[latin9]{inputenc}
\usepackage[a4paper]{geometry}
\usepackage{color}
\usepackage{amsmath}
\usepackage{graphicx}
\usepackage{amssymb}

\makeatletter
\@ifundefined{textcolor}{}
{%
 \definecolor{BLACK}{gray}{0}
 \definecolor{WHITE}{gray}{1}
 \definecolor{RED}{rgb}{1,0,0}
 \definecolor{GREEN}{rgb}{0,1,0}
 \definecolor{BLUE}{rgb}{0,0,1}
 \definecolor{CYAN}{cmyk}{1,0,0,0}
 \definecolor{MAGENTA}{cmyk}{0,1,0,0}
 \definecolor{YELLOW}{cmyk}{0,0,1,0}
 }

\makeatother

\usepackage{babel}

\begin{document}

\title{The Efimov effect in lithium 6\\
\emph{L'effet d'Efimov dans le lithium 6}}

\author{Pascal Naidon and Masahito Ueda}

\affiliation{ERATO Macroscopic Quantum Project, JST, Tokyo, 113-0033 Japan}

\affiliation{Department of Physics, University of Tokyo, 7-3-1 Hongo, Bunkyo-ku,
Tokyo 113-0033, Japan}
\begin{abstract}
We analyse the recent experiments investigating the low-energy physics
of three lithium 6 atoms in three different internal states with resonant
two-body scattering lengths. All observed features are qualitatively
consistent with the expected Efimov effect, \emph{i.e.} the effective
universal three-body attraction that arises for large values of the
scattering lengths. However, we find that a quantitative description
at negative energy requires non-universal two- and three-body corrections
due to presently unknown behaviour at short distance. An attempt to
implement these corrections is made through energy-dependent parameters
that are fitted to the experimental data.\\

\emph{Nous analysons les récentes expériences concernant la physique
à basse énergie de trois atomes de lithium 6 dans des états internes
différents avec des longueurs de diffusion résonantes. Toutes les
observations s'expliquent qualitativement par l'effet d'Efimov, c'est-à-dire
l'attraction universelle effective à trois corps qui survient aux
grandes valeurs des longueurs de diffusion. En revanche, il apparaît
qu'une compréhension quantitative nécessite des corrections non-universelles
à deux et trois corps provenant du comportement inconnu à courte distance.
Nous proposons d'implémenter ces corrections par le biais de paramètres
dépendant de l'énergie et ajustés à l'expérience.}
\end{abstract}
\maketitle
In 1970, Vitaly Efimov predicted a peculiar effect~\cite{rf:efimov},
a universal effective attraction occurring between three quantum particles
with short-range interactions, whenever the two-body interactions
are resonant, that is to say their respective two-body scattering
lengths are much larger than the typical short range of those interactions.
. Interestingly, this attraction can bind the three particles to
form trimers irrespective of the sign of the scattering lengths, \emph{i.e.}
whether the two-body interaction supports a two-body bound state of
similar energy or not. In particular, there is a domain of low energies
and large scattering lengths where the three-body physics is dominated
by the Efimov attraction. Because this attraction scales as the inverse
square of the mean distance between the three particles, as does the
kinetic energy, the physics in that region is invariant under discrete
scale transformations. As a result, for an infinite scattering length,
an infinite set of trimers accumulate at zero-energy, and all these
trimers in that region have the same structure described by only a
few parameters. For this reason, it is referred to as the (three-body)
universal region~\cite{rf:ferlaino}.

Within the last year, the Efimov effect has received a wealth of experimental
confirmations from the ultracold atom community, for several kinds
of atomic species~\cite{rf:williams,rf:barontini,rf:zaccanti,rf:pollack,rf:gross,rf:lompe1,rf:nakajima}.
Evidence of Efimov trimer states appeared as strong enhancement of
inelastic collisions due to the presence of an Efimov trimer just
below the collisional threshold. These successes were made possible
thanks to the existence of magnetic-field induced Fano-Feshbach resonances~\cite{rf:chin},
which enable experimentalists to adjust the two-body scattering length
to very large values. The case of lithium 6 atoms \cite{rf:ottenstein,rf:huckans,rf:williams,rf:lompe1,rf:nakajima}
is particularly striking for several reasons. First of all, unlike
the species used in other experiments, lithium 6 atoms are fermions.
Since the Efimov attraction does not occur for identical fermions
because of the Pauli exclusion principle, they have to be prepared
in three distinguishable states 1, 2 and 3. As a result, unlike identical
bosons, this system is described by three different scattering lengths,
and features three dimers with different binding energies. Fortunately
there exists a Feshbach resonance for each pair in the same range
of magnetic field, enabling one to have large scattering lengths for
all pairs at the same time and the Efimov effect to manifest itself
in various ways~\cite{rf:dincao-1}. In that region, the atoms can
pair to form three different kinds of dimers (12), (23), and (31).
This creates the possibility of chemical exchange reactions in the
universal region, such as (12)+3$\to$(13)+2 where the energies intersect
\cite{rf:knoop2}. Finally, because those dimers are comparatively
stable, it is possible to study the connection between dimers and
trimers at negative energy. The three-component lithium 6 gas is the
first system where a dimer and an atom could be associated into one
of the trimers, enabling direct trimer spectroscopy~\cite{rf:lompe2}.

In this paper, we examine in detail how to theoretically reproduce
the experimental observations of the Efimov effect in lithium 6. First,
we describe the two-body physics, explaining how to model the Feshbach
resonance occuring for each pair of atoms. Then, we explain how to
deal with the three-body problem, making use of some approximations.
We finally apply these models to analyse the experimental observations.

\section{Two-body physics: the Feshbach resonances}

A lithium 6 atom has a nuclear spin $i=1$ and a valence electron
spin $s=1/2$, and their coupling leads to 6 different hyperfine states
in the electronic ground state. These states can be separated by applying
a magnetic field, using the Zeeman effect. They are labelled from
1 to 6 in order of increasing energy $E_{i=1,\dots,6}$. Since lithium
6 atoms are fermions, we consider all the possible antisymmetrised
pairs $\{i,j\}$ of these states as a basis of diatomic channels.
The interaction between two atoms $A$ and $B$ depends on the arrangement
of the total electronic spin $\vec{S}=\vec{s}_{A}+\vec{s}_{B}$ into
singlet ($S=0$) or triplet ($S=1$) states. It can be written as:\begin{equation}
\hat{V}(r)=V_{0}(r)\hat{P}_{0}+V_{1}(r)\hat{P}_{1},\label{eq:InteractionPotential}\end{equation}
where $r$ is the distance between the two atoms, $\hat{P}_{S=0,1}$
are the projectors onto the singlet and triplet states, and $V_{S=0,1}(r)$
are the respective singlet and triplet potentials. Because the interaction
is non-diagonal in the diatomic channel basis, it leads to a set of
coupled Schrödinger equations describing the relative motion of a
pair of atoms. As a result, a bare scattering state in one channel
can be coupled to a bare bound state in another channel. Since these
states have different magnetic moments, it is possible to change their
energy difference by applying an external magnetic field. As these
states get closer in energy, the scattering length of the dressed
states (solution of the full equations including the coupling) becomes
larger. At some point, it diverges and changes sign. That phenomenon
is a so-called Fano-Feshbach resonance \cite{rf:chin}. Thanks to
the existence of such resonances, it is possible to make the scattering
lengths very large by controlling the magnetic field.

To quantify the effects of the resonances used in the lithum 6 experiments,
we solved the coupled-channel equations using singlet and triplet
potentials for lithium. The potentials were obtained from the combination
of ab-initio calculations and adjustments to previous experimental
data from Ref.~\cite{rf:bartenstein}, and are currently the most
accurate potentials for lithium 6. They were kindly provided to us
by Paul~S.~Julienne and Eite Tiesinga. From the solutions, we extract
important quantities characterising the interaction. For scattering
states of an atom pair $\{i,j\}$ at some positive energy $E=\frac{\hbar^{2}}{m}p^{2}$
($m$ is the mass of lithium 6) just above the channel threshold $E_{i}+E_{j}$,
the component in the channel $\{i,j\}$ (the so-called \emph{open
channel}) has the following asymptotic form\begin{equation}
\psi_{ij}^{E}(\vec{r})\propto\frac{\sin pr}{pr}-a_{ij}(p)\frac{\cos pr}{r}\label{eq:Asymptotic}\end{equation}
which defines the energy-dependent scattering length $a_{ij}(p)$.
This quantity is directly related to the more familiar scattering
amplitude $f_{ij}(p)=-1/(ip+1/a_{ij}(p))$. Similarly, for states
of negative energy $E=-\frac{\hbar^{2}}{m}\kappa^{2}$ below the channel
threshold (physical bound states as well as unphysical states), the
asymptotic form becomes\begin{equation}
\psi_{ij}^{E}(\vec{r})\propto\frac{\sinh\kappa r}{\kappa r}-a_{ij}(i\kappa)\frac{\cosh\kappa r}{r}\label{eq:Asymptotic2}\end{equation}
which extends the definition of $a_{ij}(p)$ to negative energy. The
only physical states are the bound states which decay exponentially
at large distance. Therefore they must satisfy the condition \[
a_{ij}(i\kappa)=1/\kappa,\]
whose solutions $\kappa_{ij}$ correspond to the discrete spectrum
of binding energies $E_{ij}=\frac{\hbar^{2}\kappa_{ij}^{2}}{m}$.

At low energy, the energy-dependent scattering length can be expanded
as\begin{equation}
\frac{1}{a_{ij}(p)}=\frac{1}{a_{ij}}-\frac{1}{2}r_{e,ij}p^{2}+\dots\label{eq:EnergyExpansion}\end{equation}
from which we can extract the zero-energy scattering length $a_{ij}$
and the effective range $r_{e,ij}$. Note that bound states with small
binding energies must satisfy the universal two-body property $\kappa_{ij}\sim1/a_{ij}$.

\begin{figure}
\includegraphics[bb=10bp 40bp 426bp 393bp,clip]{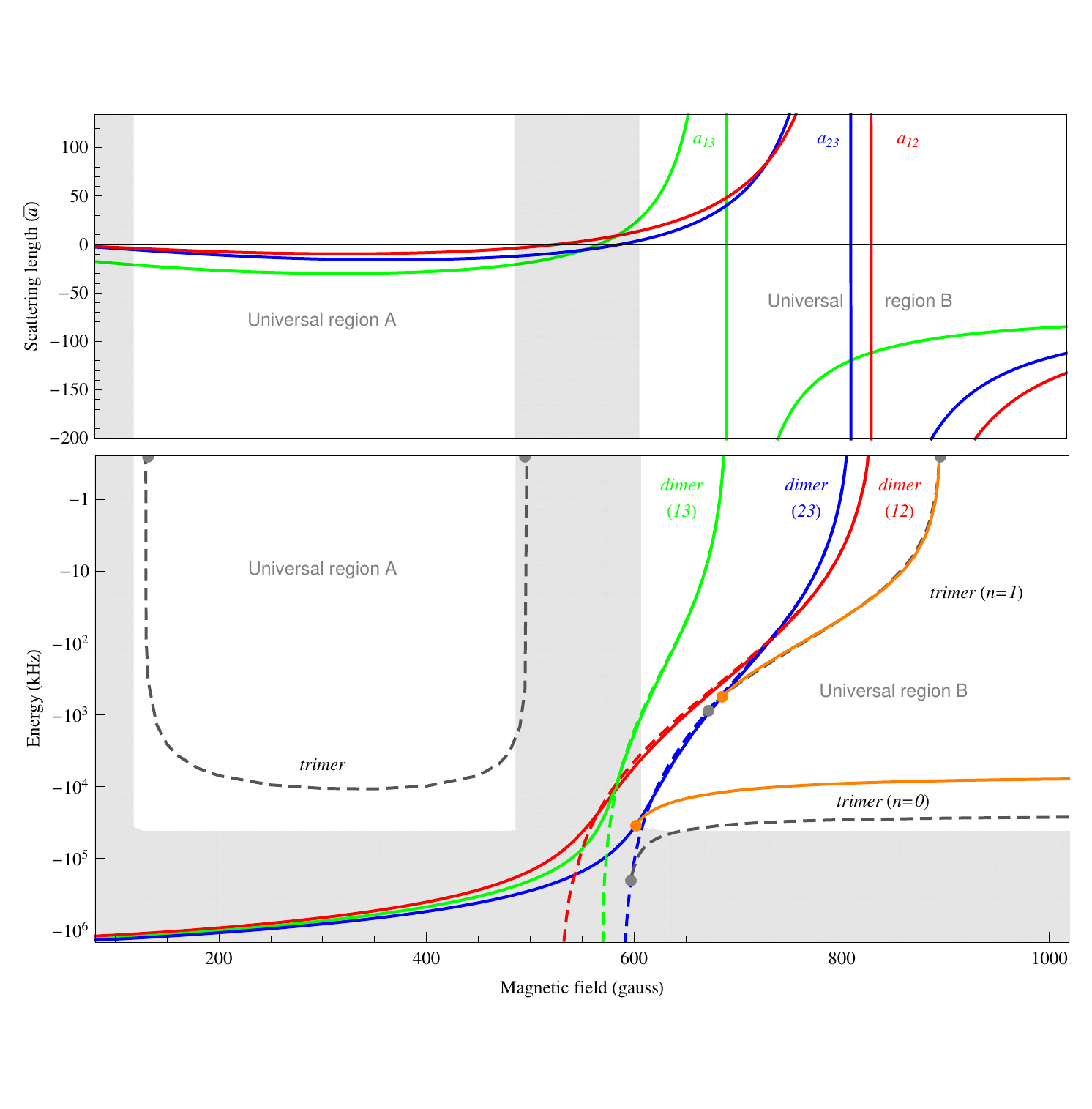}

\caption{\label{fig:ScatteringSpectrum}Top panel: scattering lengths for each
pair of atoms as a function of magnetic field. Bottom panel: two-body
and three-body energy spectra, as a function of magnetic field. The
solid coloured curves correspond to the dimers' energies, and the
dashed coloured curves are their universal limit $-\hbar^{2}/ma_{ij}^{2}$.
The dashed grey curves correspond to the trimers' energies based on
the universal theory (with $\vert\Lambda\vert=1.165\,\bar{a}^{-1}$).
The orange curves correspond to the single-channel contact model with
both two-body and three-body non-universal corrections implemented
through an energy dependence of $a_{ij}$ and $\Lambda$. The white
areas are the universal regions where all scattering lengths are larger
than $3\bar{a}$ and all energies are smaller than $\hbar^{2}/m(3\bar{a})^{2}$.}

\end{figure}

\begin{figure}
\includegraphics{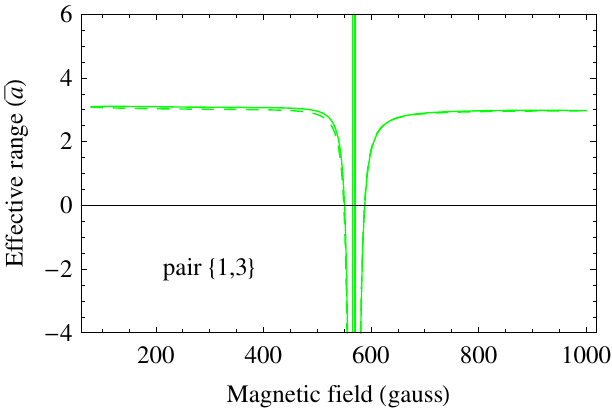}\includegraphics[bb=15bp 0bp 176bp 119bp,clip]{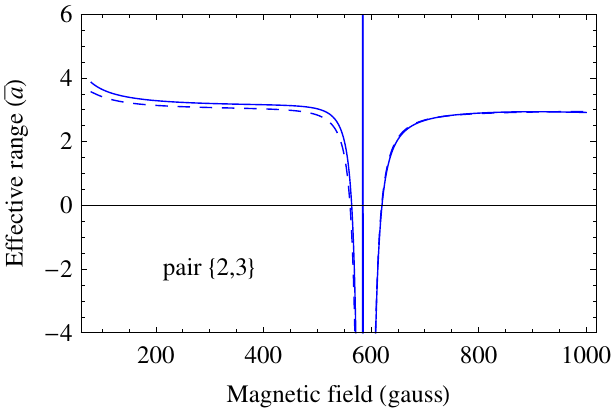}\includegraphics[bb=15bp 0bp 176bp 119bp,clip]{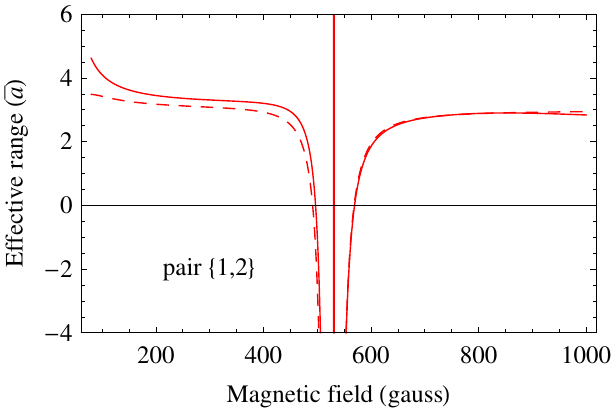}\caption{\label{fig:Effectiveranges}Effective range of the interaction for
each pair of atoms, as a function of magnetic field. Solid curves
are calculated from our multichannel two-body model. Dashed curves
are obtained from two-channel model in Eq.~(\ref{eq:ScatteringLengthModel}).}

\end{figure}

The various scattering lengths, effective ranges and binding energies
of the last diatomic bound states are represented in Figs.~\ref{fig:ScatteringSpectrum}
and \ref{fig:Effectiveranges} as a function of magnetic field for
the different atom pairs \{1,2\}, \{2,3\}, and \{1,3\} relevant to
the experiments. According to the universal two-body property $\kappa_{ij}\sim1/a_{ij}$,
a bound state appears at each resonance point where the scattering
length $a_{ij}$ becomes infinite, and its binding energy increases
as $\hbar^{2}/ma_{ij}^{2}$ on the side of positive values of the
scattering length. This universal binding energy, only valid for large
positive scattering lengths, has been represented as dashed lines
in Fig.~\ref{fig:ScatteringSpectrum}. By comparing the scattering
lengths to the typical range of the potentials, the van der Waals
length $\bar{a}$~\cite{rf:chin}, and the binding energies to the
van der Waals energy $\hbar^{2}/m\bar{a}^{2}$, we can define two
different universal regions A and B, which are indicated in Fig.~\ref{fig:ScatteringSpectrum}.
For lithium 6, $\bar{a}\approx1.5815$~nm.

For convenience, in the rest of the paper we will note $a_{jk}$,
$\kappa_{jk}$, $E_{jk}$, etc, as $a_{i}$, $\kappa_{i}$, $E_{i}$
where $\{i,j,k\}=\{1,2,3\}$.

\section{Three-body physics: approximation schemes}

It would be rather difficult to solve the problem of three atoms using
the two-body coupled potentials described in the previous section.
Moreover, on top of these two-body interactions, there is also a three-body
interaction which is presently unknown. However, since we are considering
the low-energy properties of the system, a precise knowledge of those
interactions and an exact solution of the three-body problem is not
necessary. At low energy, the wave function has a large de Broglie
wavelength and is affected by the small-scale details of the interactions
only through some phase shifts which can be set by choosing appropriate
boundary conditions at short distance~\cite{rf:efimov}. Alternatively,
one may replace the interactions by pseudo-potentials with the same
low-energy effects (causing the same phase shifts), or construct a
low-energy effective field theory~\cite{rf:braaten} with a few parameters
(corresponding again to these phase shifts). All these approaches
are essentially equivalent, and are intended to simplify the problem
by reducing the number of coordinates, and invoking just a few parameters
instead of the complicated knowledge of the two- and three-body interactions.
Here we will use the pseudopotential method. We will consider two
different pseudopotentials: a single-channel contact pseudopotential,
and a two-channel separable pseudopotential of finite range.

\subsection{Single-channel contact pseudopotential}

\begin{figure}
\includegraphics[bb=10bp 0bp 346bp 259bp,clip,scale=0.5]{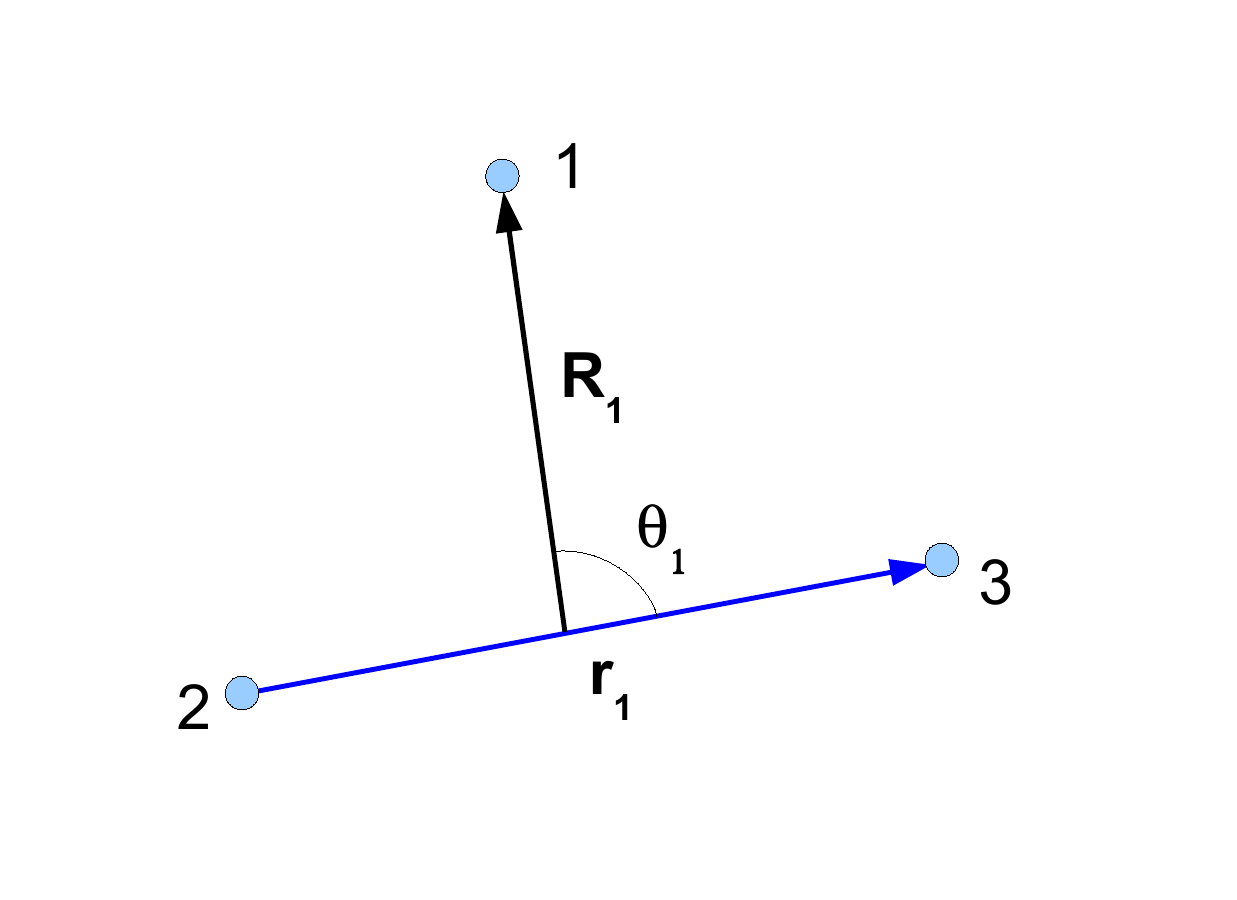}\includegraphics[bb=10bp 0bp 346bp 259bp,clip,scale=0.5]{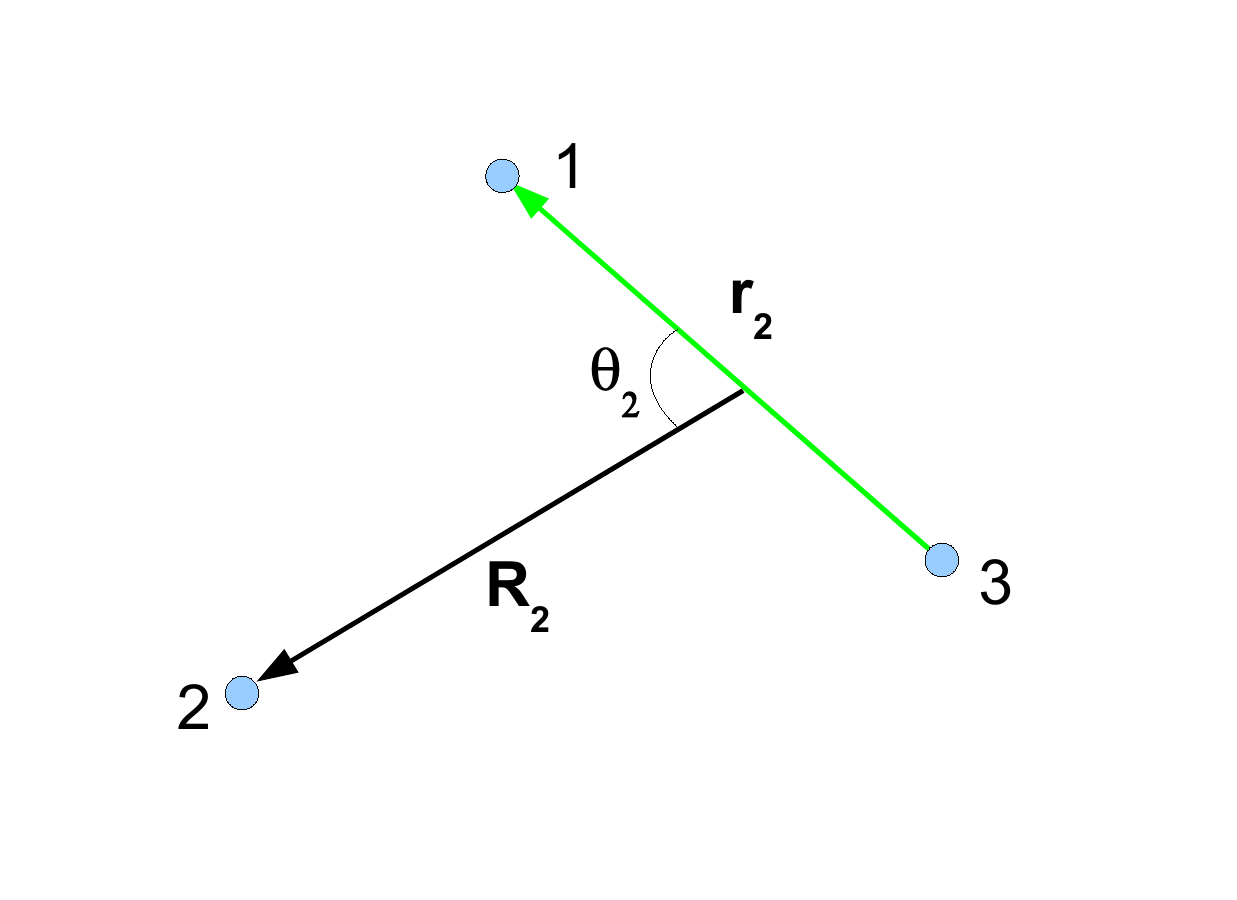}\includegraphics[bb=10bp 0bp 346bp 259bp,scale=0.5]{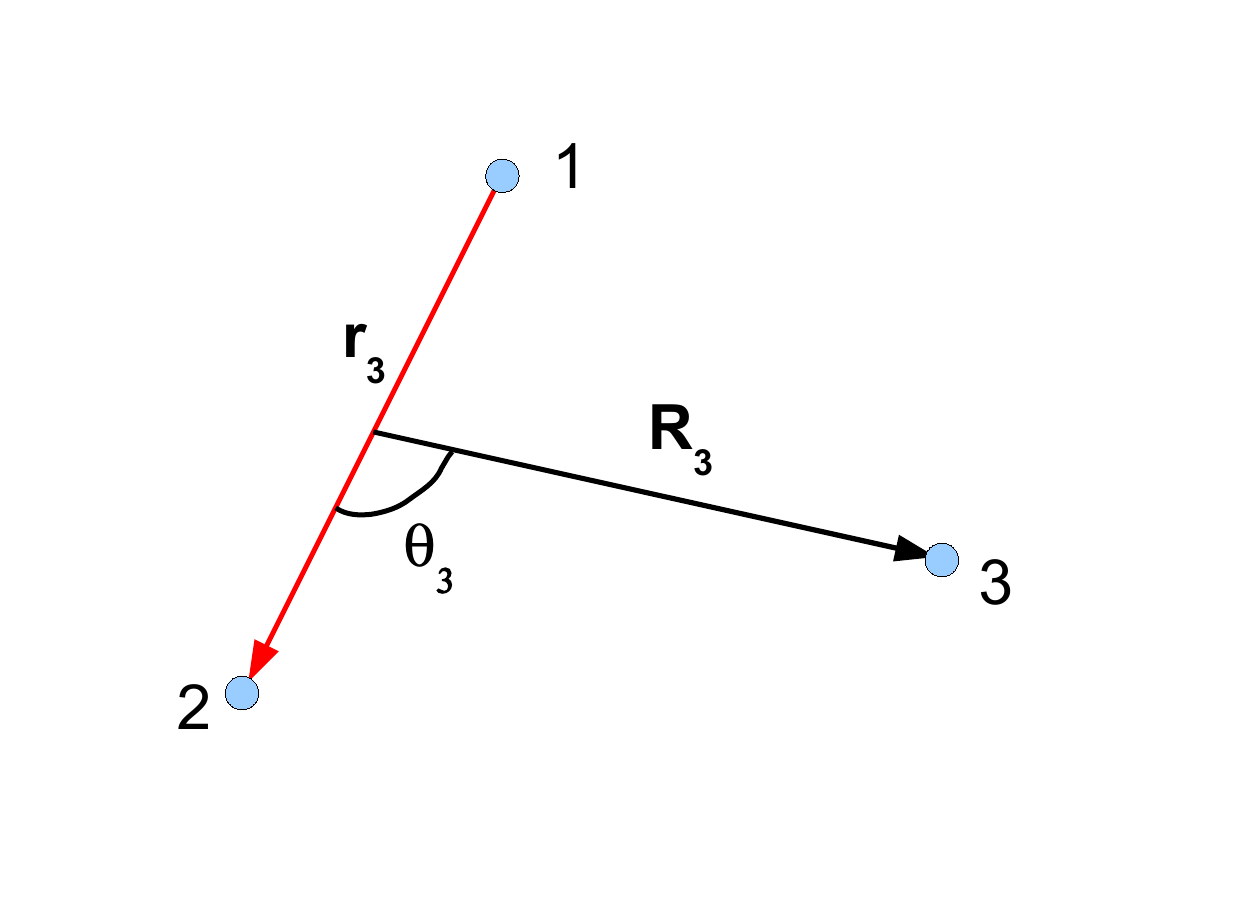}

\caption{\label{fig:Jacobi}Jacobi coordinate systems ($\vec{R}_{i},\vec{r}_{i}$).
In Fourier space, we have conjugate wave vectors ($\vec{P}_{i},\vec{p}_{i}$).}

\end{figure}

Let us start by considering a single-channel pseudopotential as a
substitute for the real interaction. The asymptotic behaviour of the
two-body scattering state in the open channel in Eqs.~(\ref{eq:Asymptotic})
and (\ref{eq:Asymptotic2}) can be reproduced by considering a noninteracting
wave $\psi$ with energy $\frac{\hbar^{2}p^{2}}{m}$ and imposing
the Bethe-Peierls boundary condition $\psi\underset{r\to0}{\propto}\frac{1}{r}-\frac{1}{a(p)}$,
or equivalently $(r\psi(r))^{\prime}/(r\psi(r))\xrightarrow[r\to0]{}-1/a(p)$
which fixes the logarithmic derivative of $\psi$ at short distance
$r$. This condition is equivalent to a Fermi-Huang-Yang contact pseudopotential
$\hat{V}$ defined by: \begin{equation}
\langle\vec{r}\vert\hat{V}\vert\psi\rangle=\frac{4\pi\hbar^{2}a(p)}{m}\delta^{3}(\vec{r})\frac{\partial}{\partial r}(r\psi(\vec{r}))\quad\mbox{with }p^{2}=\lim_{r\to0}\frac{-\nabla_{r}^{2}\psi(r)}{\psi(r)}\label{eq:PseudoPotential}\end{equation}
By constructing such a pseudopotential $\hat{V}_{i}(\vec{r}_{i})$
for each pair of atoms $\{jk\}$, the 3-body Schrödinger equation
reads\begin{equation}
\left[-\frac{\hbar^{2}}{m}\left(\frac{3}{4}\nabla_{R}^{2}+\nabla_{r}^{2}\right)-E+\hat{V}_{1}(\vec{r}_{1})+\hat{V}_{2}(\vec{r}_{2})+\hat{V}_{3}(\vec{r}_{3})\right]\Psi(\vec{R},\vec{r})=0,\label{eq:ThreeBodyEquation}\end{equation}
where we have chosen a particular set of Jacobi coordinates $(\vec{R},\vec{r})=(\vec{R}_{1},\vec{r}_{1})$
to represent the relative configuration of three atoms in states 1,
2, and 3 - see Fig.~\ref{fig:Jacobi}. The pseudopotentials ensure
that in all three Jacobi coordinate systems, the three-body wavefunction
$\Psi$ should locally have the following form~\cite{rf:Petrov}:
\begin{equation}
\Psi(\vec{R}_{i},\vec{r}_{i})\underset{r_{i}\to0}{=}\left(\frac{1}{r_{i}}-\frac{1}{a_{i}(p_{i})}\right)\chi_{i}(\vec{R}_{i})+O(r_{i}),\label{eq:BethePeierlsForm}\end{equation}

\[
\mbox{with }p_{i}^{2}=\lim_{r_{i}\to0}\frac{-\nabla_{\vec{r}_{i}}^{2}\Psi}{\Psi}=\frac{mE}{\hbar^{2}}+\frac{3}{4}\frac{\nabla_{\vec{R}_{i}}^{2}\chi_{i}(\vec{R}_{i})}{\chi_{i}(\vec{R}_{i})},\]
which is consistent with the Bethe-Peierls boundary condition for
each pair of atoms. This introduces the three quantities $\chi_{i}(\vec{R}_{i})$.
From Eq.~(\ref{eq:ThreeBodyEquation}), we obtain\begin{equation}
\left(-\frac{3}{4}\nabla_{R}^{2}-\nabla_{r}^{2}-\frac{mE}{\hbar^{2}}\right)\Psi(\vec{R},\vec{r})=4\pi\sum_{i=1,2,3}\chi_{i}(\vec{R}_{i})\delta^{3}(\vec{r}_{i}).\label{eq:SchrodingerEqWithPseudopotential}\end{equation}
The solution of this equation in Fourier space is

\begin{equation}
\tilde{\Psi}(\vec{P},\vec{p})=\tilde{\Psi}_{0}(\vec{P},\vec{p})+\frac{4\pi}{\frac{3}{4}P^{2}+p^{2}-mE/\hbar^{2}+i\varepsilon}\sum_{i=1,2,3}\tilde{\chi}_{i}(\vec{P}_{i})\label{eq:Psi}\end{equation}
where the term $\tilde{\Psi}_{0}$ is a solution of the homogeneous
(free) 3-body equation. It physically corresponds to an incident wave
of 3 free atoms $(2\pi)^{6}\delta^{3}(\vec{P}-\vec{P}_{0})\delta^{3}(\vec{p}-\vec{p}_{0})$
in the case of positive energy $E=\frac{\hbar^{2}}{m}(\frac{3}{4}P_{0}^{2}+p_{0}^{2})$,
or it must be taken to be zero in the case of states with negative
energy $E$, \emph{i.e.} states $\Psi$ with at least two bound atoms.
From Eq.~(\ref{eq:Psi}) and the definition of $\chi_{i}$, it follows
that the $\chi_{i}$ must satisfy a set of 3 coupled equations known
as the Skorniakov~-~Ter-Martirosian equations~\cite{rf:skorniakov},

\begin{equation}
\left(\frac{-1}{a_{i}(i\gamma_{P})}+\gamma_{P}\right)\tilde{\chi}_{i}(P)-\frac{1}{\pi}\int_{0}^{\Lambda}\!\!\! dq\frac{q}{P}\ln\frac{P^{2}+q^{2}+Pq-mE/\hbar^{2}}{P^{2}+q^{2}-Pq-mE/\hbar^{2}}(\tilde{\chi}_{j}(q)+\tilde{\chi}_{k}(q))=\int\!\!\!\frac{d^{3}\vec{p}}{(2\pi)^{3}}\tilde{\Psi}_{0}(\vec{P},\vec{p})\label{eq:STM}\end{equation}
with the relative momentum $\gamma_{P}=\sqrt{\frac{3}{4}P^{2}-\frac{mE}{\hbar^{2}}}$.
The derivation is given in Appendix A. If we take the upper bound
$\Lambda$ of the integral in the left-hand side of Eq.~(\ref{eq:STM})
to be infinite, the equations admit several solutions, as was noted
by G.~V.~Skorniakov and discussed by G.~S.~Danilov~\cite{rf:danilov}.
This is because the replacement of real interactions by two-body contact
interactions is in general not sufficient to have a well-defined 3-body
problem. In special situations of two-body interactions with a large
negative effective range, a unique solution is determined by the equations~\cite{rf:Petrov}.
Otherwise, an extra boundary condition on the wave function at very
short distance where all three atoms are close to one another is necessary.
Cutting off the integral at some momentum $\Lambda$ \cite{rf:Kharchenko}
precisely provides such a condition, as shown in Appendix B.

In his original papers, Vitaly Efimov considered an energy-independent
scattering length $a(p)=a$. As we mentioned before, at the two-body
level this approximation is valid only in the universal region of
large scattering lengths (compared to the range of the interactions).
Hence, Eqs. (\ref{eq:STM}) with an energy-independent scattering
length are equivalent to the \emph{universal theory} originally proposed
by Vitaly Efimov to investigate the universal properties of 3-body
systems with large scattering lengths. Note that the energy-dependence
of $a(k)$ brings finite-range corrections to the universal theory,
including the effective range correction~\cite{rf:efimov2} as can
be seen from Eq. (\ref{eq:EnergyExpansion}).

\subsection{Coupled-channel separable pseudopotential}

Although the previous approach does take into account finite-range
corrections, it may still look oversimplified because it neglects
the coupled-channel nature of the real atomic interactions near Feschbach
resonances. To account for this, one may substitute the real interaction
by an effective two-channel interaction $\hat{U}$ that reproduce
most features of the original Feshbach resonances~\cite{rf:lee,rf:jonalasinio}.
It can be written as \begin{equation}
\hat{U}=\hat{U}_{o}\vert o\rangle\langle o\vert+\hat{U}_{c}\vert c\rangle\langle c\vert+\hat{U}_{oc}\vert c\rangle\langle o\vert+\hat{U}_{co}\vert o\rangle\langle c\vert\label{eq:CoupledChannelPotential}\end{equation}
where $\vert o\rangle$ and $\vert c\rangle$ correspond to open and
closed channels, respectively. 

Let us first look at two atoms interacting with this effective potential.
The two-body Hamiltonian is $\hat{K}+\hat{U}$, where $\hat{K}$ is
the relative kinetic energy. To model a resonance with a single bound
state $\phi$ of energy $E_{b}$ in the closed channel, we take the
closed-channel Hamiltonian $\hat{K}+\hat{U}_{c}$ to be $E_{b}\vert\phi\rangle\langle\phi\vert$
and the channel-coupling term as $\hat{U}_{oc}=\hat{U}_{co}^{\dagger}=W\vert\phi\rangle\langle\phi\vert$.
A two-body state $\vert\psi\rangle$ can be written as a superposition
of open and closed-channel components $\vert\psi\rangle=\int\frac{d^{3}\vec{p}}{(2\pi)^{3}}A(\vec{p})\vert\vec{p}\rangle\vert o\rangle+B\vert\phi\rangle\vert c\rangle$,
where $\vert\vec{p}\rangle$ is the plane wave for two atoms with
relative momentum $\vec{p}$, and the Schrödinger equation $(\hat{K}+\hat{U}-E)\vert\psi\rangle=0$
leads to:\begin{eqnarray}
(E_{b}-E)B & +2\int\frac{d^{3}\vec{q}}{(2\pi)^{3}}A(\vec{q})W\phi^{*}(\vec{q}) & =0,\label{eq:CoupledEq1}\\
(\frac{\hbar^{2}k^{2}}{m}-E)A(\vec{p}) & +\int\frac{d^{3}\vec{q}}{(2\pi)^{3}}A(\vec{q})U_{o}(\vec{q},\vec{p})+B\, W\phi(\vec{p}) & =0.\label{eq:CoupledEq2}\end{eqnarray}

By assuming that the open-channel potential is separable \cite{rf:lee,rf:jonalasinio2008,rf:werner2009,rf:jonalasinio},
$U_{o}(\vec{q},\vec{p})=\frac{4\pi\hbar^{2}}{m}\lambda\phi^{*}(\vec{q})\phi(\vec{p})$,
we can group the last two terms in the second equation, which greatly
simplifies the problem. This arbitrary choice is permitted since the
small-scale details of the interaction are unimportant as long as
they correctly reproduce the low-energy physics, as argued before.
For the same reason, we can arbitrarily choose the functional form
of $\phi(p)$. Following \cite{rf:lee,rf:jonalasinio}, we choose
for convenience a Gaussian form $\phi(p)=e^{-\frac{1}{2}(bp)^{2}}$
with range $b$.

Eliminating $B$ in Eq. (\ref{eq:CoupledEq2}) and solving at positive
energy $E=\frac{\hbar^{2}p_{0}^{2}}{m}$, one obtains:\begin{equation}
A(\vec{p})=A_{0}(\vec{p})-\frac{T(p_{0},p)}{\frac{\hbar^{2}p^{2}}{m}-E-i\varepsilon}\label{eq:TwoBodyAmplitude}\end{equation}
where $A_{0}(\vec{p})=(2\pi)^{3}\delta^{3}(\vec{p}-\vec{p}_{0})$
corresponds to an incident plane wave of momentum $\vec{p}_{0}$,
and we introduced the two-body $T-$matrix element \begin{equation}
T(p_{0},p)=\left(\frac{4\pi\hbar^{2}}{m}\lambda-\frac{2\vert\Lambda\vert^{2}}{E_{b}-E}\right)\left(\int\frac{d^{3}\vec{q}}{(2\pi)^{3}}\phi^{*}(q)A(q)\right)\phi(p).\label{eq:TMatrix}\end{equation}
Solving for $T(p_{0},p)$ self-consistently using the last two formul\ae,
we deduce that the scattering length $a(p)=-\left[\left(\frac{m}{4\pi\hbar^{2}}T(p,p)\right)^{-1}+ip\right]^{-1}$
takes the form~\cite{rf:werner2009}

\begin{equation}
a(p)=\left\{ \left[\left(\lambda-\frac{\alpha}{E_{b}-E}\right)^{-1}+\frac{1}{\sqrt{\pi}b}\right]e^{(pb)^{2}}+p\mbox{Erfi}(pb)\right\} ^{-1}\label{eq:ScatteringLengthModel}\end{equation}
where $\alpha=\frac{m}{2\pi\hbar^{2}}\vert\Lambda\vert^{2}$, and
Erfi is a real function related to the standard error function Erf
by $\mbox{Erfi}(z)=\mbox{Erf}(iz)/i$. By adjusting the parameters
$\lambda$, $\alpha$, $E_{b}$ and $b$, we can construct for each
pair of atoms $\{j,k\}$ a pseudopotential $\hat{U}_{i}$ which reproduces
the energy-dependent scattering length $a_{i}(p)$ of the real interaction.

We now consider three atoms 1, 2, 3 interacting through these pseudopotentials
$\hat{U}_{1}$, $\hat{U}_{2}$ and $\hat{U}_{3}$. The 3-body wave
function can be written as:\begin{equation}
\vert\Psi\rangle=\int\frac{d^{3}\vec{P}}{(2\pi)^{3}}\frac{d^{3}\vec{p}}{(2\pi)^{3}}A(\vec{P},\vec{p})\vert\vec{P}\rangle_{1}\vert\vec{p}\rangle_{23}+\sum_{(i,j,k)=(1,2,3)}\int\frac{d^{3}\vec{P}}{(2\pi)^{3}}B_{i}(\vec{P})\vert\vec{P}\rangle_{i}\vert\phi\rangle_{jk}\label{eq:ThreeBodyState}\end{equation}
where $\vert\vec{P}\rangle_{i}$ is the plane wave with momentum $\vec{P}$
for the relative motion between atom $i$ and the centre of mass of
the pair of atoms $j$ and $k$, $\vert\vec{p}\rangle_{jk}$ is the
plane wave of momentum $\vec{p}$ for the relative motion between
atoms $j$ and $k$, and $\vert\phi\rangle_{jk}$ is the closed-channel
bound state for the pair of atoms $j$ and $k$. Defining the quantity
\begin{equation}
\tilde{\beta}_{i}(q)=-\left(\lambda_{i}-\frac{\Lambda_{i}}{E_{i}+\frac{3}{4}\frac{\hbar^{2}Q_{k}^{2}}{m}-E+i\varepsilon}\right)\int\frac{d^{3}\vec{p}}{(2\pi)^{3}}\phi^{*}(p_{i})A(Q_{i},p_{i}),\label{eq:DefinitionOfBeta}\end{equation}
we obtain an equation similar to Eq.~(\ref{eq:Psi}) for the open-channel
component $A$,$ $\begin{equation}
A(\vec{P},\vec{p})=A_{0}(\vec{P},\vec{p})+\frac{4\pi}{\frac{3}{4}P^{2}+p^{2}-mE/\hbar^{2}+i\varepsilon}\sum_{i=1,2,3}\tilde{\beta}_{i}(\vec{P}_{i})\phi(\vec{p}_{i}),\label{eq:OpenChannelComponentEquation}\end{equation}
where $A_{0}$ is an incident plane wave $(2\pi)^{6}\delta^{3}(\vec{P}-\vec{P}_{0})\delta^{3}(\vec{p}-\vec{p}_{0})$
for positive energy $E=\frac{\hbar^{2}}{m}(\frac{3}{4}P_{0}^{2}+p_{0}^{2})$,
or zero for negative energy. The $\tilde{\beta}_{i}$ then satisfy
the generalised Skorniakov - Ter-Martirosian coupled equations, \begin{multline}
\vert\phi(i\gamma_{P})\vert^{2}\left(\frac{-1}{a_{i}(i\gamma_{P})}+\gamma_{P}\right)\!\tilde{\beta}_{i}(P)-\frac{1}{\pi}\int_{0}^{\Lambda}\!\!\! q^{2}dq\!\int_{-1}^{1}\!\!\!\! du\frac{\phi^{*}(\sqrt{q^{2}+\frac{P^{2}}{4}+Pqu})\phi(\sqrt{\frac{q^{2}}{4}+P^{2}+Pqu})}{P^{2}+q^{2}-Pq-mE/\hbar^{2}}(\tilde{\beta}_{j}(q)+\tilde{\beta}_{k}(q))\\
=\int\!\!\!\frac{d^{3}\vec{p}}{(2\pi)^{3}}\phi^{*}(\vec{p})A_{0}(\vec{P},\vec{p})\label{eq:GSTM-2}\end{multline}

Note that these equations are very similar to Eqs.~(\ref{eq:STM}).
In particular, all the two-body physics is contained in the energy-dependent
scattering length $a_{i}$, except for the terms $\phi$. When $\phi\to1$,
\emph{i.e.} when the range $b$ of the interaction goes to zero, we
formally retrieve the single-channel contact interaction equations
(\ref{eq:STM}). This indicates that apart from the terms $\phi$
there is little difference between the two approaches. Here, the presence
of $\phi$ with nonzero range $b$ effectively cuts off the integral
at high momenta, so that we can safely take $\Lambda\to\infty$. In
fact, the range $b$ of the pseudo-potential plays the role of $\Lambda$
in Eqs.~(\ref{eq:STM}), $i.e.$ it characterises the 3-body behaviour
at short distance.

It might seem unreasonable to choose $b$, which is determined by
the two-body interaction only, in order to characterise a three-body
property. In general, the short-range three-body behaviour should
also depend on a three-body interaction between the atoms. To be consistent,
a three-body interaction should be added in the problem. Another approach,
taken in Ref.~\cite{rf:jonalasinio}, is to regard $b$ not as a
parameter describing the real two-body interaction (for example its
effective range), but as a free two-body parameter that is adjusted
to set the combined effects of the real two-body and three-body interactions.

\subsection{Models}

To analyse the experimental results, we will consider three different
three-body models:
\begin{itemize}
\item The \emph{universal model} is given by Eqs.~(\ref{eq:STM}) with
energy-independent scattering lengths $a_{i}$ that are obtained from
the two-body calculation of section I, and $\Lambda$ is a free parameter. 
\item The \emph{single-channel contact model} is given by Eqs.~(\ref{eq:STM})
with energy-dependent scattering lengths $a_{i}(p)$ given by Eqs.~(\ref{eq:ScatteringLengthModel}),
where we set the parameters $\lambda_{i}$ $\alpha_{i}$, $E_{b,i}$,
and $b$ so as to reproduce the two-body quantities calculated in
Section I, namely $a_{i}$, $r_{e,i}$ and $\kappa_{i}$, as a function
of magnetic field. The reason why we use this two-body model instead
of the real $a_{i}(p)$ is that it is numerically intractable to calculate
the real $a_{i}(p)$ at negative energy due to the divergence of sinh
and cosh at large distance in Eq.~(\ref{eq:Asymptotic2}). Nevertheless,
this simple two-channel model should be very close to the real $a_{i}(p)$
in the low-energy domain that we are interested in.
\item The \emph{two-channel separable model} is given by Eqs.~(\ref{eq:GSTM-2})
with the scattering lengths $a_{i}(p)$ given again by Eqs.~(\ref{eq:ScatteringLengthModel}).
\end{itemize}

\section{Analysis of the experiments}

There has been two sets of experiments. The first one \cite{rf:ottenstein,rf:huckans,rf:williams}
consisted in measuring the three-body recombination rate in a gas
of atoms equally distributed in the lowest three spin states. Three-body
recombination is the process where three colliding atoms recombine
to form a combination of states with lower internal energy, such as
a deep dimer and a free atom. At the Feshbach resonance locations,
the three-body recombination is strongly enhanced by the very large
scattering length. Away from these points, other peaks were found
and attributed to recombination enhancement by the presence of Efimov
trimers at zero energy. Indeed, whenever a trimer state exists just
below the collisional threshold of three atoms, the three atoms resonate,
which increases their probability to be close together and recombine.
In the second set of experiments \cite{rf:lompe1,rf:nakajima}, a
gas of dimers and atoms was prepared, and inelastic collisions by
relaxation to deeper dimer states were observed. The relaxation rate
was also found to be enhanced at two magnetic field values due to
the presence of two Efimov trimers just below the collision threshold.
Recently, one of the two trimers' energy was directly observed by
association spectroscopy \cite{rf:lompe2,rf:nakajima2}. These experiments
thus provided some partial information about the spectrum of Efimov
trimers at both zero and and negative energies. The general spectrum
based on these results is given in Fig.~\ref{fig:ScatteringSpectrum}.

\subsection{Experiments at zero energy (3-body recombination)}

When three atoms in different states collide and recombine, the density
$n_{i}$ of atoms in each state $i$ decreases according to the rate
equation:\begin{equation}
\dot{n}_{i}=-K_{\mbox{rec}}n_{i}n_{j}n_{k},\label{eq:RateEquation}\end{equation}
where $K_{\mbox{rec}}$ is the recombination coefficient. By measuring
the variation of the number of atoms, and taking into account other
kinds of loss, it is possible to extract the recombination coefficient.
The clearest evidence of the enhancement of this coefficient by the
presence of an Efimov trimer at zero-energy is the peak found around
895~G by Williams et al.~\cite{rf:williams}. It is indeed located
in a region of very large scattering lengths where the universal theory
should be valid

To calculate the recombination coefficient, we proceed as follows.
We distinguish between two types of dimers: dimers which are included
in the theory (through the solutions of $1/a(i\kappa)=\kappa$), referred
to as \emph{shallow dimers}, and dimers which are not included in
the theory, referred to as \emph{deep dimers}. Recombination to shallow
dimer $(jk)$ appears in the 3-body wave function as an outgoing wave
between dimer $(jk)$ and atom $i$. This means that $\tilde{\chi}_{i}$
or $\tilde{\beta}_{i}$ can be written as:\begin{equation}
\tilde{\chi}_{i}(P)=\sqrt{\mathcal{N}_{i}}\frac{4\pi F_{i}(P)}{P^{2}-Q_{i}^{2}-i\varepsilon}\quad;\quad\tilde{\beta}_{i}(P)=\sqrt{\mathcal{N}_{i}^{\prime}}\frac{4\pi F_{i}(P)}{P^{2}-Q_{i}^{2}-i\varepsilon},\label{eq:DefinitionOfScatteringAmplitudes}\end{equation}
where $Q_{i}=\sqrt{\frac{4}{3}(\frac{m}{\hbar^{2}}E+\kappa_{i}^{2})}$
is the relative momentum between dimer $(jk)$ and atom $i$, and
$\mathcal{N}_{i}=\left(\int\frac{d^{3}p}{(2\pi)^{3}}\left|\frac{1}{p^{2}+\kappa_{i}^{2}}\right|^{2}\right)^{-1}=\frac{\kappa_{i}}{2\pi}$
and $\mathcal{N}_{i}^{\prime}=\left(\int\frac{d^{3}p}{(2\pi)^{3}}\left|\frac{\phi(p)}{p^{2}+\kappa_{i}^{2}}\right|^{2}\right)^{-1}\mathcal{P}_{o}$
are factors ensuring that the dimer wavefunction is properly normalised
to unity, or the probability $\mathcal{P}_{o}$ to be in the open
channel, respectively. The recombination coefficient $K_{\mbox{rec}}^{i}$
to shallow dimers $(jk)$ is then obtained by calculating the flux
of that outgoing wave~\cite{rf:Petrov,rf:jonalasinio2008}:\begin{equation}
K_{\mbox{rec}}^{i}=\frac{3h}{m}Q_{i}\vert F_{i}(Q_{i})\vert^{2}.\label{eq:PartialRecombinationCoefficient}\end{equation}

\begin{figure}
\includegraphics{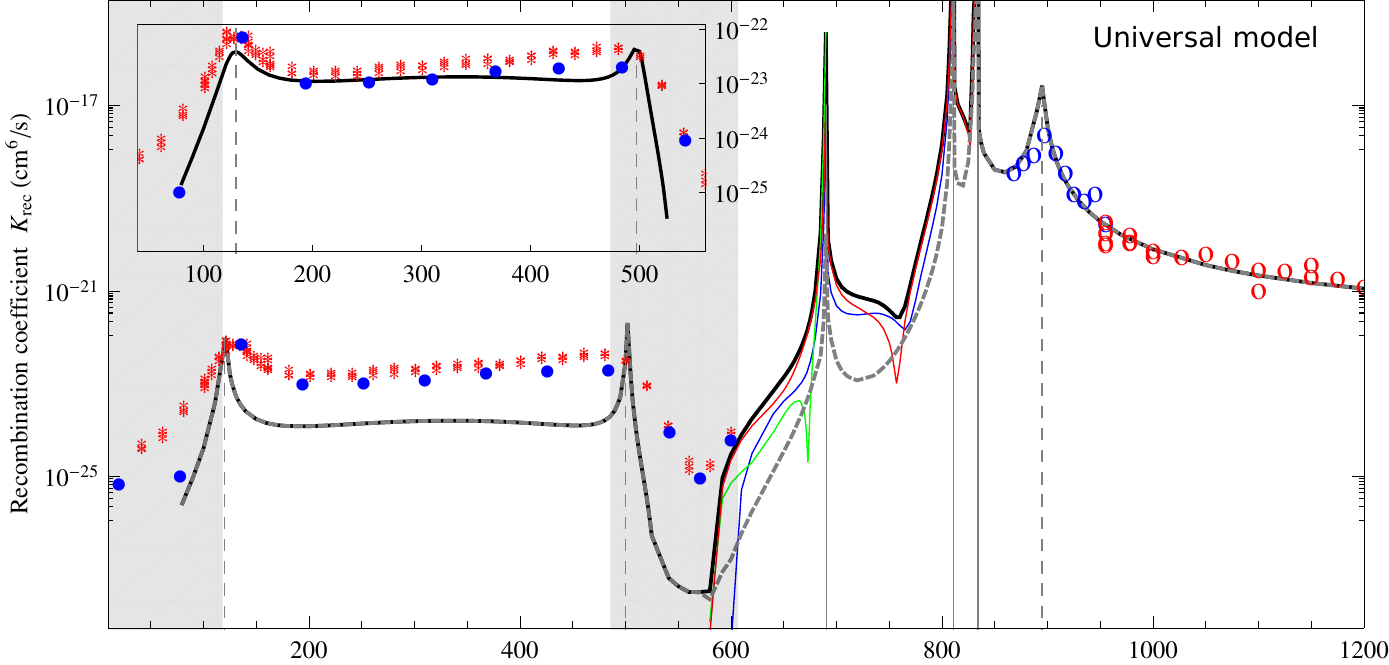}

\includegraphics{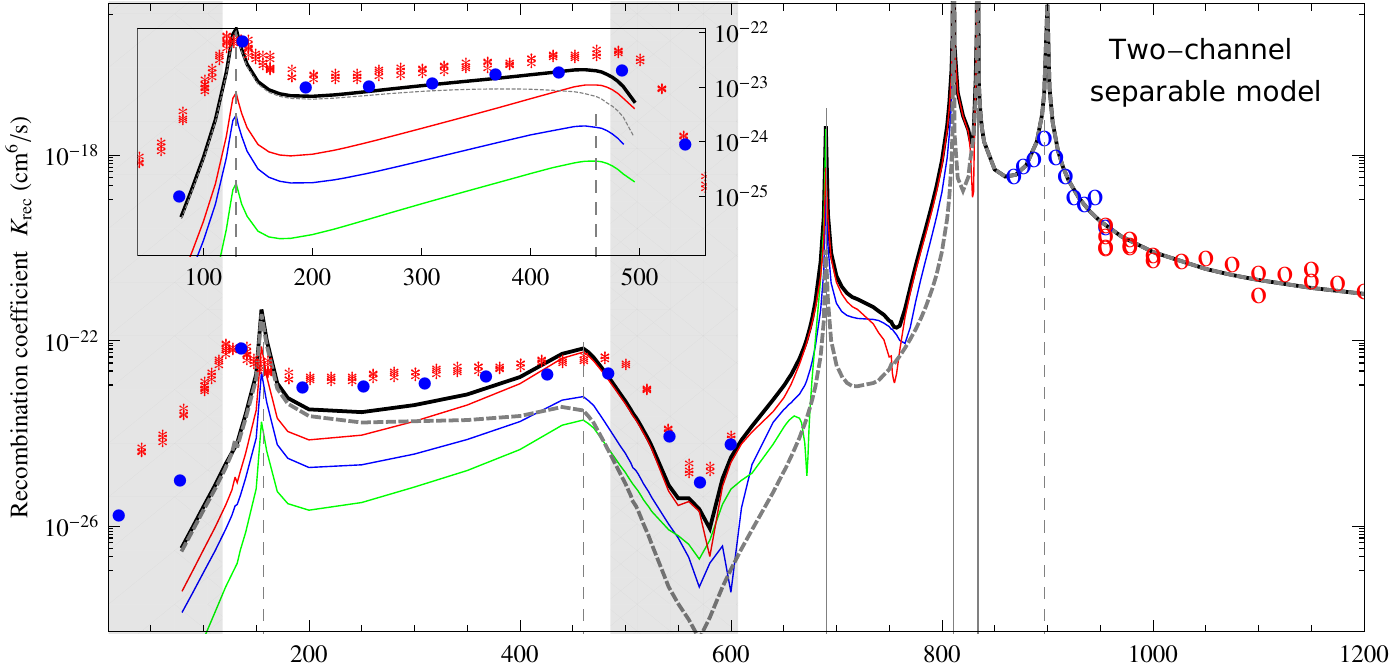}

\caption{\label{fig:Recombination}Recombination coefficient $K_{\mbox{rec}}$
as a function of magnetic field. The stars, dots and circles indicate
the experimental measurements of Refs.~\cite{rf:ottenstein}, \cite{rf:huckans},
and \cite{rf:williams}, respectively. Top panel: calculations from
the universal model for $\vert\Lambda\vert=$ $1.165\,\bar{a}^{-1}$
and $\eta=$ 0.016. The inset shows results in universal region A
for $\vert\Lambda\vert=$ $1.076\,\bar{a}^{-1}$ and $\eta=$ 0.115.
Bottom panel: calculations from the two-channel separable model for
$b=1.31\,\bar{a}$ and $\xi=0.08$. The inset shows results in universal
region A for $b=0.75\,\bar{a}$ and $\xi=0.5$. In both panels, thick
black curves show the total recombination coefficient, dashed grey
curves show the recombination coefficient to deep dimers, and coloured
curves show the recombination to shallow dimers: (13) green, (23)
blue, and (12) red. }

\end{figure}

Recombination to deep dimers occurs at distances on the order of the
deep dimers' size, typically given by the range of the interactions.
Therefore, the coefficient for recombination to deep dimers can be
estimated by calculating the probabibility of finding the three atoms
within that range \cite{rf:PetrovSalomon,rf:WernerCastin2006}:\begin{equation}
K_{\mbox{rec}}^{\mbox{(deep)}}=\xi\frac{\hbar}{m\mathcal{R}_{0}^{2}}\int_{\sqrt{\frac{4}{3}R^{2}+r^{2}}<\mathcal{R}_{0}}d^{3}\vec{R}d^{3}\vec{r}\vert\Psi(\vec{R},\vec{r})\vert^{2}\label{eq:DeepDimerRecombinationCoefficient}\end{equation}
The results are not very sensitive to the precise choice of the range
$\mathcal{R}_{0}$, and the constant factor $\xi$ is expected to
be on the order of unity. Typically, if $\mathcal{R}_{0}$ varies
by 10\%, the rate $K_{\mbox{rec}}^{\mbox{(deep)}}$ changes by 6\%.
As a typical size, we chose $\mathcal{R}_{0}=\bar{a}$. The total
recombination coefficient is $K_{\mbox{rec}}=\sum_{i}K_{\mbox{rec}}^{i}+K_{\mbox{rec}}^{\mbox{deep}}$.
Alternatively, in the contact interaction model, one can set a complex
three-body parameter $\Lambda=\vert\Lambda\vert e^{i\eta}$, where
$\eta>0$ phenomenologically reproduces short-distance losses due
to recombination to deep dimers. In that case, the total recombination
coefficient at zero energy is given by (see Appendix C) \begin{equation}
K_{\mbox{rec}}=\frac{4h}{m}\mbox{Im}\sum_{i}\tilde{\chi}_{i}(0).\label{eq:TotalLossRateCoefficient}\end{equation}
This method also has the advantage of taking into account the broadening
effect of the loss strength $\eta$ on the recombination profile as
a function of magnetic field. We checked that the two methods give
similar results. The calculated recombination coefficients for different
models are represented in Fig.~\ref{fig:Recombination}.

In the universal model, we have to adjust $\vert\Lambda\vert$ to
$1.165\,\bar{a}^{-1}$ and $\eta$ to 0.016 ($\xi\approx0.8$) in
order to reproduce the 895~G peak. This peaks correspond to the appearance
of an Efimov trimer at the 3-body collisional threshold (zero energy)
as represented in Fig.~\ref{fig:ScatteringSpectrum}.Then we obtain
the variation of the recombination coefficient around 895~G in the
whole universal region B, as first calculated by Eric Braaten et al.
\cite{rf:braaten3}. We extend the calculation to low magnetic field
in universal region A where earlier measurements of recombination
were performed \cite{rf:ottenstein,rf:huckans}. Interestingly, the
same three-body parameter roughly reproduces the variation of the
recombination coefficient: a plateau with two peaks on both ends.
However the two peaks are not exactly at the right locations. This
is not unexpected, since different universal regions separated by
zero crossings of the scattering lengths are known to have different
3-body phases in general~\cite{rf:dincao}. Changing the value of
$\vert\Lambda\vert$ to 1.076~$\bar{a}^{-1}$, and $\eta$ to 0.115
gives a fair agreement between the measured recombination and the
universal model - see the inset in the top panel of Fig.~\ref{fig:Recombination}.
Note that the two peaks correspond again to zero-energy crossings
of a trimer in Fig.~\ref{fig:ScatteringSpectrum}. This was in fact
the first indication of the underlying Efimov physics in lithium~6~\cite{rf:braaten2,rf:naidon1,rf:floerchinger}.
However the calculated recombination coefficient shows two marked
peaks, while in the experimental data the left peak is much more pronounced
than the right one. Variations of the loss parameter $\eta$ with
magnetic field were subsequently proposed to improve the agreement
of the universal model with the data \cite{rf:wenz,rf:rittenhouse}.
The reason is that in the universal model the least-bound dimers cease
to exist at low magnetic field because the scattering length becomes
negative. Since in reality those dimers are still present, the loss
parameter $\eta$ effectively accounts for transitions to these dimers.
As their energy varies with magnetic field, it was inferred that $\eta$
should also change with magnetic field~\cite{rf:rittenhouse}. 

In the single-channel contact model and the two-channel separable
model, the least-bound dimers are explicitly included over the full
range of magnetic field. For binding energies larger than the typical
van der Waals energy associated with the range the interaction, corresponding
to magnetic field smaller than 600~G, their description becomes unrealistic.
However their energy remains accurate, and they still provide a simple
model for the recombination mechanism. In the single-channel contact
model, we adjust $\vert\Lambda\vert$ to $1.455\,\bar{a}^{-1}$ and
$\eta=0.0033$ ($\xi\approx0.8$) to reproduce the 895~G peak. Note
that because the scattering lengths are now energy-dependent, the
values of $\vert\Lambda\vert$ and $\eta$ have been altered in order
to obtain the same physical situation. This is because the choice
of the high-energy (\emph{i.e.} short-distance) behaviour of the two-body
interactions affects the choice of the three-body phase. The calculated
recombination rate in the universal region B is nearly identical to
the universal model. It is however numerically difficult to extend
the calculation to the universal region A, presumably because in that
region the dimers' binding momenta $\kappa_{i}$ are close or even
exceed the cutoff momentum $\Lambda$. This problem does not occur
for the two-channel separable model.

In the two-channel separable model, we first adjusted $b$ to 1.31~$\bar{a}$
to match the effective range of the two-body interaction - see Fig.~\ref{fig:Effectiveranges}.
Suprisingly, this choice perfectly reproduces the location of the
peak at 895~G. Again, the results in universal region B are very
similar to those of the universal model. In universal region A, however,
the results are different and significantly off the observed peak
locations. Changing $b$ to 0.75~$\bar{a}$ (thus using a wrong effective
range, while preserving the correct scattering lengths and dimer binding
energies) we can effectively change the 3-body phase and get better
agreement - see bottom panel inset in Fig.~\ref{fig:Recombination}.
These numerical results also seem consistent with the semi-analytical
approach of Ref.~\cite{rf:rittenhouse} based on the universal theory
and a magnetic-field-dependent $\eta$.

From all these results, we conclude that the 3-body physics of lithium
6 at zero energy is essentially consistent with the universal theory.

\subsection{Experiments at negative energy}

\begin{figure}
\includegraphics{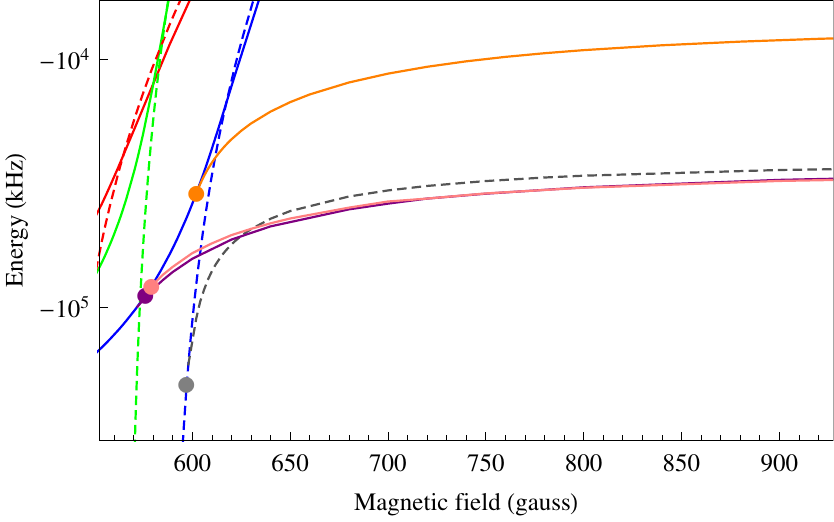}\includegraphics{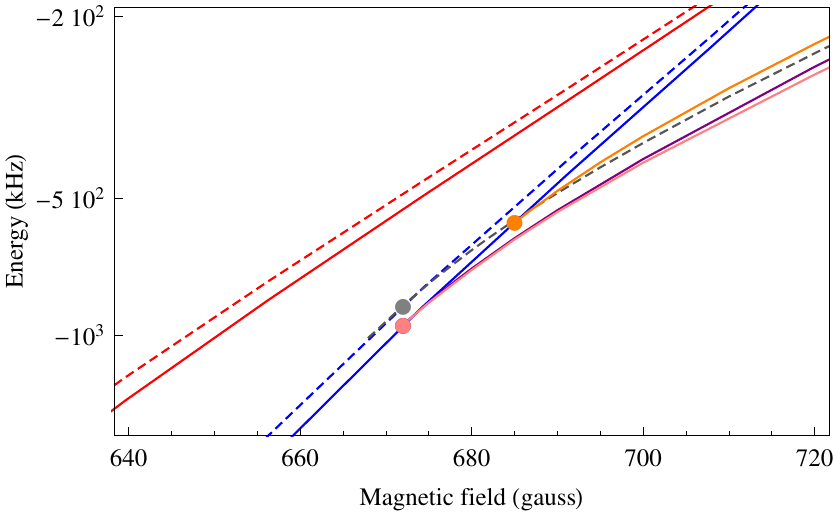}\caption{\label{fig:Trimers}Trimers and dimers connection regions. Left: ground-state
Efimov trimer. Right: first excited-state Efimov trimer. Dimer curves
are the same as in Fig.~\ref{fig:ScatteringSpectrum}. Trimer curves
are obtained for different models: universal model (dashed grey),
single-channel contact model (purple), two-channel separable model
(pink), and single-channel model with the energy-dependent parameter
$\Lambda$ given in Fig.~\ref{fig:ThreeBodyParameter} (orange).
All models feature a zero-energy resonance at 895~G.}

\end{figure}

From the previous analysis, it is possible to predict the energy spectrum
of trimers using the previously adjusted parameters. The spectrum
based on the universal model, first predicted in Refs.~\cite{rf:naidon1,rf:floerchinger,rf:braaten3},
is shown in Fig.~\ref{fig:ScatteringSpectrum}. In universal region
A, it predicts a trimer state which connects to no dimer but dissociates
into three atoms at two magnetic field values corresponding to the
two 3-body recombination peaks. In universal region B, it predicts
the existence of two trimers which connect to the dimer (23) at 598~G
and 672~G. In fact, the first meeting point is outside the universal
region where the dimer binding energy clearly deviates from the universal
formula $\hbar^{2}/ma^{2}$. Therefore, it was expected in Ref.~\cite{rf:braaten3}
to be unreliable. On the other hand, the 672~G prediction is right
in the universal region and was thought to be reliable. It turns out
that the dimer energy still has a small but appreciable deviation
from universality at that magnetic field, due to two-body finite-range
corrections. However, computing the trimer energy with the single-channel
contact or two-channel separable model (both of which include those
two-body finite range corrections) leads to a similar trimer energy
curve which again meets the dimer curve at around 672 G - see Fig.~\ref{fig:Trimers}.
In other words, the two-body finite range corrections shift both the
dimer and trimer energies, but do not modify the magnetic field of
their meeting point.

\subsubsection{Atom-dimer relaxation}

To check those predictions, experimentalists prepared a mixture of
dimers (23) and atoms 1, and observed the rate of relaxation to deep
dimers as atoms and dimers collide. The density $n_{1}$ of atoms
in state $1$ therefore decreases according to the rate equation:\begin{equation}
\dot{n}_{1}=-K_{\mbox{rel}}n_{1}n_{23},\end{equation}
where $n_{23}$ is the density of dimers (23) and $K_{\mbox{rel}}$
is the relaxation coefficient. Calculating the relaxation coefficient
from the theory is similar to the case of the recombination coefficient.
For the case of dimer (23) colliding with atom 1 at energy $E-E_{12}=\frac{3}{4}\frac{\hbar^{2}Q_{1}^{2}}{m}$,
the quantities $\tilde{\chi}_{i}$ can be written as:\begin{equation}
\chi_{i}(P)=\sqrt{\mathcal{N}_{i}}\left(\delta_{i1}(2\pi)^{3}\delta^{3}(\vec{P}-\vec{Q}_{1})+\frac{4\pi f_{i}(P)}{P^{2}-Q_{i}^{2}-i\varepsilon}\right)\end{equation}
corresponding to an incident wave and outgoing waves of amplitude
$f_{i}(Q_{i})$ and momentum $Q_{i}=\sqrt{\frac{4}{3}(\frac{m}{\hbar^{2}}E+\kappa_{i}^{2})}$.
The relaxation coefficient $K_{\mbox{rel}}^{i\ne1}$ to shallow dimers
$(jk)$ is then obtained by calculating the flux of the corresponding
outgoing wave:\begin{equation}
K_{\mbox{rel}}^{i}=\frac{4h}{m}Q_{i}\vert f_{i}(Q_{i})\vert^{2},\label{eq:RelaxationCoefficient}\end{equation}
and the total relaxation coefficient is\begin{equation}
K_{\mbox{rel}}=\frac{4h}{m}\left(\mbox{Im}f_{1}(Q_{1})-Q_{1}\vert f_{1}(Q_{1})\vert^{2}\right).\label{eq:TotalRelaxationCoefficient}\end{equation}

\begin{figure}
\includegraphics{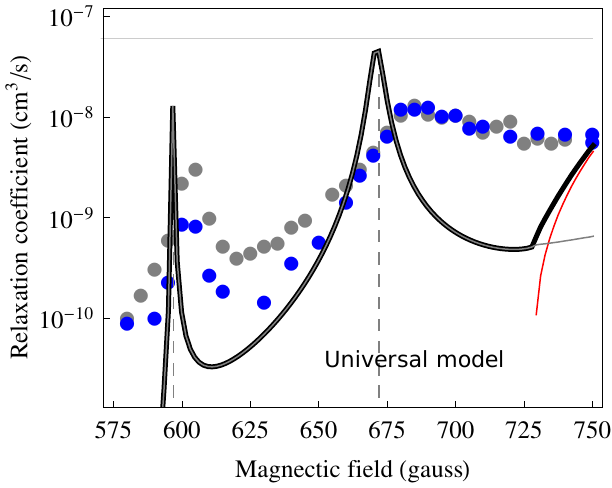}\includegraphics[bb=27bp 0bp 176bp 140bp,clip]{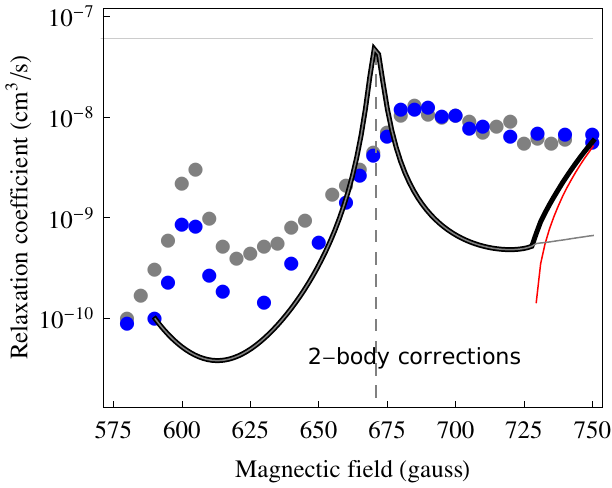}\includegraphics[bb=27bp 0bp 176bp 140bp,clip]{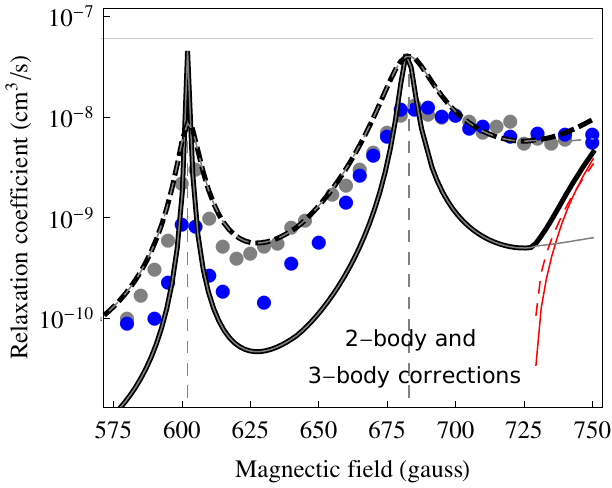}

\caption{\label{fig:Relaxation23}Relaxation coefficient for dimer 23 colliding
with atom 1 as a function of magnetic field. The grey and blue dots
indicate the measurements from Refs.~\cite{rf:lompe1} and \cite{rf:nakajima}.
Left: universal model with $\vert\Lambda\vert=1.165\,\bar{a}^{-1}$
and $\eta=0.016$ - see similar calculations in Refs.~\cite{rf:braaten3,rf:hammer}.
Middle: contact model with $\vert\Lambda\vert=1.455\,\bar{a}^{-1}$
and $\eta=0.0033$. Right: contact model with the energy-dependent
$\vert\Lambda\vert$ of Fig.~\ref{fig:ThreeBodyParameter} and $\eta=0.0033$
(dashed curves: $\eta=0.0400$). Black curves show the total relaxation
coefficient, red curves show the relaxation to dimer 12, and grey
curves show the relaxation to deep dimers. The horizontal line indicates
the unitarity limit given by the typical collisional energy ($\sim$100~nK)
in the experiments.}

\end{figure}

Similarly to 3-body recombination, whenever the magnetic field is
close to the meeting point between a dimer and trimer, a resonance
occurs which strongly enhances the relaxation coefficient. Indeed,
two peaks were observed near the expected meeting points, at 602~G
and 685~G \cite{rf:lompe1,rf:nakajima} - see Fig.~\ref{fig:Relaxation23}.
However, the significant deviation of the second peak location from
the expected value 672~G is somewhat surprising. Indeed, we checked
that if we change the 3-body parameter to obtain a peak at the measured
685~G, then the peak in the 3-recombination coefficient moves to
865~G, which seems incompatible with the measured value of 895~G
in Ref.~\cite{rf:williams}.

\begin{figure}
\includegraphics{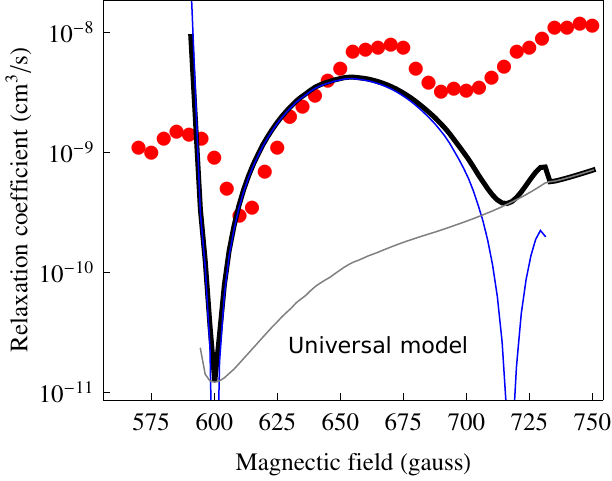}\includegraphics[bb=27bp 0bp 176bp 138bp,clip]{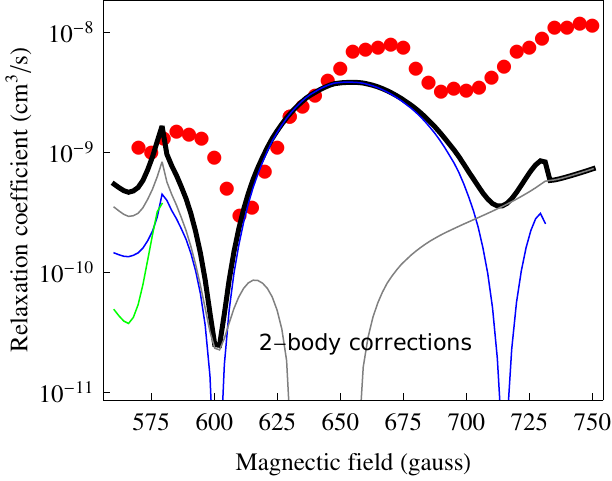}\includegraphics[bb=27bp 0bp 176bp 138bp,clip]{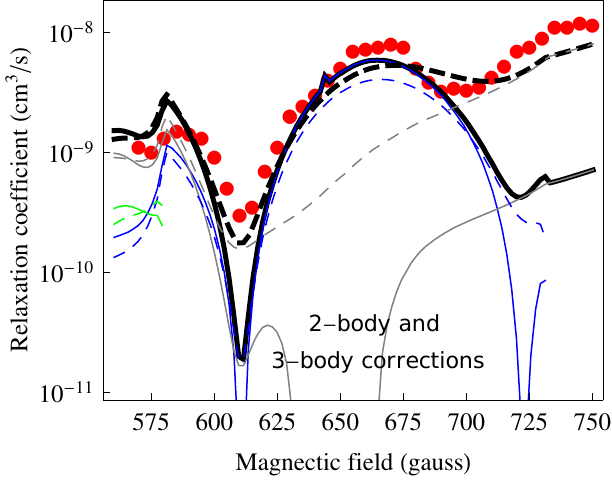}

\caption{\label{fig:Relaxation12}Relaxation coefficient for dimer 12 colliding
with atom 3 as a function of magnetic field. The red dots show the
results of the measurements from Ref.~\cite{rf:lompe1}. The same
conventions as in Fig.~\ref{fig:Relaxation23} are used.}

\end{figure}

Similar measurements of the relaxation coefficient involving a dimer~(12)
colliding with atom~3~\cite{rf:lompe1} revealed some dips in the
relaxation coefficient at 610~G and 690~G - see Fig.~\ref{fig:Relaxation12}.
These dips are expected to result from two-pathway interferences related
to the Efimov physics of the system~\cite{rf:dincao-1}. Actually
only the first dip appears as a clear signature of such interference.
The second dip, as our calculations suggest, may be due to the combined
effect of relaxation to shallow dimers and deep dimers. While both
the universal and single-channel contact models predict a dip at 600~G,
the real dip is again shifted away from the theoretical expectation. 

\begin{figure}
\includegraphics{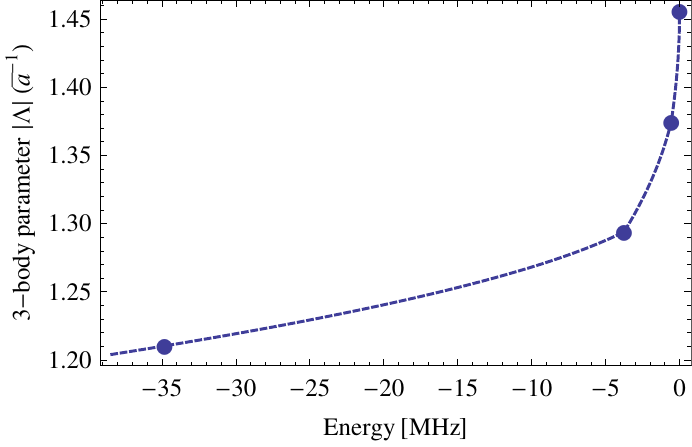}

\caption{\label{fig:ThreeBodyParameter}Energy dependence of the three-body
parameter $\Lambda$ (in units of $\bar{a}^{-1}$) of the single-channel
contact model adjusted to fit measured resonances. Each dot corresponds
to an adjustment to a peak or dip of the relaxation coefficient.}

\end{figure}

Since all non-universal two-body corrections have been taken into
account in the theoretical models, we concluded in Ref.~\cite{rf:nakajima}
that a non-universal three-body correction is also needed. In the
contact model, non-universal two-body corrections arise from the energy
dependence of the scattering lengths $a_{i}(p)$. In the same fashion,
we expect the three-body parameter $\Lambda$ to be energy-dependent
in general. By adjusting the value of $\Lambda(p)$ to fit each peak
and dip locations we obtain some insight on this energy dependence.
The result is plotted in Fig.~\ref{fig:ThreeBodyParameter} and looks
consistent with a smooth but non-linear variation with energy. By
interpolating $\Lambda(p)$ as a function of $p$, we can recalculate
all previous curves and obtain reasonable agreement with experimental
data, see Figs.~\ref{fig:ScatteringSpectrum}, \ref{fig:Relaxation23},
and \ref{fig:Relaxation12}. This adjusted non-universal 3-body model
predicts that the ground-state Efimov trimer at large magnetic field
is shifted by about 20~MHz from the universal prediction. Direct
measurement of that energy would clearly validate or invalidate our
assumption of the energy dependence of the 3-body parameter $\Lambda$.
The fact that the 3-body parameter of an effective theory is energy-dependent
to describe the ground-state trimer is not suprising, since ground
states of Efimov series are always non-universal. It is less obvious,
however, why it is also energy-dependent near the second trimer, which
should be closer to universality.

\subsubsection{Atom-dimer association spectroscopy}

\begin{figure}
\includegraphics{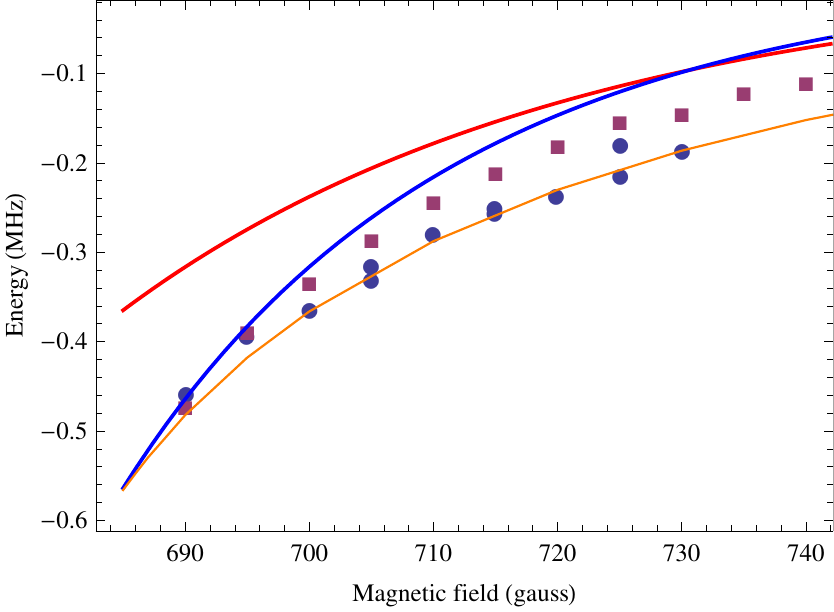}

\caption{\label{fig:Direct-measurements}Direct measurements of the second
trimer energy: dots correspond to measurements of Ref.~\cite{rf:lompe2},
which were taken at about 1 $\mu K$ , and squares show the results
of the measurements in Ref.~\cite{rf:nakajima2} which were taken
at lower temperature where thermal shifts are negligible. Using the
same conventions as Fig.~\ref{fig:ScatteringSpectrum}, thick curves
represent the dimer energies and the orange curve represents the trimer
energy obtained from the single-channel contact model with the energy-dependent
3-body parameter shown in Fig.~\ref{fig:ThreeBodyParameter}.}

\end{figure}

Very recently, the binding energy of the excited Efimov trimer was
directly observed by association spectroscopy~\cite{rf:lompe2,rf:nakajima2}.
The experimentalists prepared a mixture of dimers and atoms, and applied
a radio-frequency field to induce a transition to the trimer state.
The preliminary results \cite{rf:lompe2} reported excellent agreement
with our own theoretical prediction \cite{rf:nakajima} of the binding
energy based on the single-channel contact model with the energy-dependent
three-body parameter $\Lambda(p)$ - see Fig.~\ref{fig:Direct-measurements}.
However, it turns out that the thermal shift of the association peaks
is non-negligible in this experiment, and was not taken into account
in the reported measurements of the binding energies. Correcting for
these shifts makes the measurements deviate significantly from the
original prediction. Worse still, to account for the deviation, the
parameter $\Lambda(p)$ should be adjusted in a way which breaks its
expected smooth variation with energy. This problem is the subject
of a separate study~\cite{rf:nakajima2}.

\section{Conclusion}

In this paper we provided an overview of the various experimental
results concerning the Efimov physics in a three-component lithium
6 system. Using different theoretical models, we found that the Efimov
features measured at nearly zero energy are essentially consistent
with the so-called Efimov scenario based on the universal theory developed
in Vitaly Efimov's original papers. However, the features measured
at negative energy, although qualitatively consistent with the Efimov
scenario, show some significant deviations from universal theory,
as well as theories which fully take into account non-universal two-body
corrections. To account for this, we introduced phenomenological non-universal
3-body corrections through an energy dependence of the three-body
parameter adjusted the experimental data. While most measurements
can be explained by this \emph{ad hoc} parametrization, it does not
seem to be consistent with the very recent measurements of a trimer's
binding energy, and further studies are needed to fully undertand
the physics at negative energy. The two-body model which the present
study is based on is thought to be very accurate. However, even minor
inaccuracies at the two-body level has been shown to be important
in the three-body physics of other systems\cite{rf:gross2}, and therefore
deserve further investigation in our system as well.

\subsubsection*{Appendix A}

Here, we give the derivation of Eq.~(\ref{eq:STM}). From Eq.~(\ref{eq:Psi}),
we perform the inverse Fourier transformation with respect to the
second variable:\begin{equation}
\tilde{\Psi}(\vec{P},\vec{r})=\tilde{\Psi}_{0}(\vec{P},\vec{r})+\left[\frac{e^{-\gamma_{P}\vert\vec{r}\vert}}{\vert\vec{r}\vert}\tilde{\chi}_{1}(\vec{P})+4\pi\int\frac{\tilde{\chi}_{2}(\vec{P}_{2})+\tilde{\chi}_{3}(\vec{P}_{3})}{\frac{3}{4}P^{2}+p^{2}-mE/\hbar^{2}+i\varepsilon}e^{i\vec{p}\cdot\vec{r}}\frac{d^{3}\vec{p}}{(2\pi)^{3}}\right].\end{equation}
Then, applying the operator $\lim_{r\to0}\frac{\partial}{\partial r}(r\cdot)$,
and using the definition of $\tilde{\chi}_{1}$, we find:\begin{equation}
-\frac{1}{a_{1}(i\gamma_{P})}\tilde{\chi}_{1}(\vec{P})=\left[\tilde{\Psi}_{0}(\vec{P},\vec{r})\right]_{r\to0}+\left[-\gamma_{P}\tilde{\chi}_{1}(\vec{P})+4\pi\int\frac{\tilde{\chi}_{2}(\vec{P}_{2})+\tilde{\chi}_{3}(\vec{P}_{3})}{\frac{3}{4}P^{2}+p^{2}-mE/\hbar^{2}+i\varepsilon}\frac{d^{3}\vec{p}}{(2\pi)^{3}}\right].\end{equation}
 Using $\vec{P}_{2}=-\frac{\vec{P}}{2}-\vec{p}$, and $\vec{P}_{3}=-\frac{\vec{P}}{2}+\vec{p}$,
we can change the integration variable to get\begin{eqnarray}
-\frac{1}{a_{1}(i\gamma_{P})}\tilde{\chi}_{1}(\vec{P}) & = & \left[\tilde{\Psi}_{0}(\vec{P},\vec{r})\right]_{r\to0}+\left[-\gamma_{P}\tilde{\chi}_{1}(\vec{P})+4\pi\int\frac{\tilde{\chi}_{2}(\vec{q})+\tilde{\chi}_{3}(\vec{q})}{\frac{3}{4}P^{2}+(\vec{q}+\frac{\vec{P}}{2})^{2}-mE/\hbar^{2}+i\varepsilon}\frac{d^{3}\vec{q}}{(2\pi)^{3}}\right],\end{eqnarray}
\begin{equation}
\left(\gamma_{P}-\frac{1}{a_{1}(i\gamma_{P})}\right)\tilde{\chi}_{1}(\vec{P})-4\pi\int\frac{\tilde{\chi}_{2}(\vec{q})+\tilde{\chi}_{3}(\vec{q})}{P^{2}+q^{2}+\vec{P}\cdot\vec{q}-mE/\hbar^{2}+i\varepsilon}\frac{d^{3}\vec{q}}{(2\pi)^{3}}=\left[\tilde{\Psi}_{0}(\vec{P},\vec{r})\right]_{r\to0}.\end{equation}

\begin{flushleft}
Assuming that $\tilde{\chi}_{i}(\vec{q})=\tilde{\chi}_{i}(q)$, and
writing the integration element $d^{3}\vec{q}=q^{2}dqd\varphi d(-\cos\theta)$,
with $\vec{P}\cdot\vec{q}=Pq\cos\theta$, we can perform the integration
over the angles $\varphi$ and $\theta$:\begin{eqnarray}
\left(\gamma_{P}-\frac{1}{a_{1}(i\gamma_{P})}\right)\tilde{\chi}_{1}(P)\qquad\qquad\qquad\qquad\qquad\qquad\qquad\qquad\qquad\qquad\qquad\qquad\nonumber \\
-\frac{4\pi}{8\pi^{3}}\int\left(\int_{0}^{2\pi}\!\! d\varphi\right)\left(\int_{-1}^{1}\frac{d(-\cos\theta)}{P^{2}+q^{2}-Pq\cos\theta-mE/\hbar^{2}+i\varepsilon}\right)(\tilde{\chi}_{2}(q)+\tilde{\chi}_{3}(q))q^{2}dq & = & \left[\tilde{\Psi}_{0}(\vec{P},\vec{r})\right]_{r\to0}\end{eqnarray}
\begin{equation}
\left(-\frac{1}{a_{1}(i\gamma_{P})}+\gamma_{P}\right)\tilde{\chi}_{1}(P)-\frac{1}{\pi}\int\frac{1}{Pq}\ln\frac{P^{2}+q^{2}+Pq-mE/\hbar^{2}}{P^{2}+q^{2}-Pq-mE/\hbar^{2}}(\tilde{\chi}_{2}(q)+\tilde{\chi}_{3}(q))q^{2}dq=\left[\tilde{\Psi}_{0}(\vec{P},\vec{r})\right]_{r\to0},\label{eq:OriginalSTM}\end{equation}
from which we obtain Eq.~(\ref{eq:STM}) after introducing a cutoff
$\Lambda$ to the integral.
\par\end{flushleft}

\subsubsection*{Appendix B}

Here we derive the connection between the upper bound $\Lambda$ of
the integral in Eq.~(\ref{eq:STM}) and the boundary condition at
short distance between the three particles. According to Danilov~\cite{rf:danilov}
and Minlos-Faddeev~\cite{rf:minlos}, the solutions $\tilde{\chi}_{i}(\vec{p})$
of the original Skorniakov - Ter-Martirosian equations (without upper
bound $\Lambda$) are superpositions of two linearly independent solutions,
which have the following asymptotic form:\begin{equation}
\tilde{\chi}_{i}(\vec{p})\xrightarrow[p\to\infty]{}A\frac{\sin(s_{0}\ln p)}{p^{2}}+B\frac{\cos(s_{0}\ln p)}{p^{2}}\propto\frac{\sin(s_{0}\ln\frac{p}{\Lambda_{0}})}{p^{2}},\end{equation}
where $s_{0}\approx1.00624$ is a constant. Thus, some extra condition
can determine one particular solution (up to normalisation) by fixing
the ratio $B/A=\tan(s_{0}\ln\Lambda_{0})$. The quantity $\Lambda_{0}$
is the 3-body parameter of the original Efimov theory. It is set by
imposing a boundary condition on the wave function at short distance~\cite{rf:efimov}.
Alternatively, one can cut off the integral in Eq.~(\ref{eq:STM})
at some high momentum $\Lambda$~\cite{rf:Kharchenko,rf:braaten2,rf:braaten3},
and this imposes the condition that the remaining part of the integral
be zero:\begin{equation}
\int_{\Lambda}^{\infty}\!\!\! dq\frac{q}{P}\ln\frac{P^{2}+q^{2}+Pq-mE/\hbar^{2}}{P^{2}+q^{2}-Pq-mE/\hbar^{2}}(\tilde{\chi}_{j}(q)+\tilde{\chi}_{k}(q))\approx\int_{\Lambda}^{\infty}4\frac{\sin(s_{0}\ln\frac{k}{\Lambda_{0}})}{k^{2}}dk=0\end{equation}
 This condition selects a particular solution. By evaluating the above
integral, one obtains\begin{equation}
4\frac{s_{0}\cos(s_{0}\ln\frac{\Lambda}{\Lambda_{0}})+\sin(s_{0}\ln\frac{\Lambda}{\Lambda_{0}})}{(\Lambda/\Lambda_{0})(1+s_{0}^{2})}=0\end{equation}
 which gives an explicit relation between the three-body parameter
$\Lambda_{0}$ and the imposed cutoff $\Lambda$:\begin{equation}
\Lambda_{0}=\Lambda\exp(\frac{\arctan s_{0}+\pi n}{s_{0}}),\quad\mbox{with }n\mbox{ integer.}\end{equation}

\subsubsection*{Appendix C}

The expression for the total loss rate coefficient at zero energy,
Eq.~(\ref{eq:TotalLossRateCoefficient}), can be seen as a consequence
of the optical theorem. Here we indicate a short derivation. We start
from the wave function in Eq.~(\ref{eq:Psi}) in space coordinates
at zero energy \begin{equation}
\Psi(\vec{R},\vec{r})=1+\sum_{j}\int\frac{d^{3}\vec{P}_{j}}{(2\pi)^{3}}\frac{d^{3}\vec{p}_{j}}{(2\pi)^{3}}\frac{4\pi}{\frac{3}{4}P_{j}^{2}+p_{j}^{2}+i\varepsilon}\tilde{\chi}_{j}(\vec{P}_{j})e^{i\vec{P_{j}}\cdot\vec{R_{j}}}e^{i\vec{p}_{j}\cdot\vec{r}_{j}}.\end{equation}
For large $R_{j}$, \emph{i.e.} when all three atoms are far from
each other, it can be approximated as $\Psi(\vec{R},\vec{r})=1+G_{0}(\vec{R},\vec{r})\sum_{j}\tilde{\chi}_{j}(\vec{0})$,
where\begin{equation}
G_{0}(\vec{R},\vec{r})=\int\frac{d^{3}\vec{P}_{j}}{(2\pi)^{3}}\frac{d^{3}\vec{p}_{j}}{(2\pi)^{3}}\frac{4\pi}{\frac{3}{4}P_{j}^{2}+p_{j}^{2}+i\varepsilon}e^{i\vec{P_{j}}\cdot\vec{R_{j}}}e^{i\vec{p}_{j}\cdot\vec{r}_{j}}=\frac{2\sqrt{3}}{(3\pi/2)^{2}}\frac{1}{\mathcal{R}^{4}},\end{equation}
and $\mathcal{R}=\sqrt{r^{2}+\frac{4}{3}R^{2}}$ is the hyper-radius
of the system. Calculating the flux of probability current of the
wave function through a large hypersphere $S$, we obtain \begin{eqnarray}
\frac{2\hbar}{m}\mbox{Im}\int_{S}\Psi^{*}\vec{\nabla}_{\mathcal{R}}\Psi\cdot d\vec{S} & = & \frac{2\hbar}{m}\left(\int_{S}\vec{\nabla}_{\mathcal{R}}G_{0}\cdot d\vec{S}\right)\mbox{Im}\sum_{i}\tilde{\chi}_{i}(0)\\
 & = & \frac{2\hbar}{m}\left(\left(\begin{array}{c}
\frac{\sqrt{3}}{2}\end{array}\right)^{3}\frac{dG_{0}}{d\mathcal{R}}\frac{2\pi^{3}\mathcal{R}^{5}}{\Gamma[3]}\right)\mbox{Im}\sum_{i}\tilde{\chi}_{i}(0)\\
 & = & -\frac{4h}{m}\mbox{Im}\sum_{i}\tilde{\chi}_{i}(0).\label{eq:LargeDistanceFlux}\end{eqnarray}
This incoming flux should balance the outgoing fluxes in other sectors
where two atoms are recombined into a shallow dimer {[}Eq.~(\ref{eq:PartialRecombinationCoefficient}){]},
as well as the loss induced by the imaginary part of the three-body
parameter $\Lambda$, which phenomenologically describes recombination
to deep dimers. Therefore, Eq.~(\ref{eq:LargeDistanceFlux}) represents
the total recombination coefficient $K_{\mbox{rec}}$.

\end{document}